\documentclass[pdflatex,sn-nature]{sn-jnl}

\usepackage{graphicx}                          % 插图
\usepackage{amsmath,amssymb,amsfonts}          % 数学公式
\usepackage{amsthm}
\usepackage{upgreek}
\usepackage{mathrsfs}
\usepackage[title]{appendix}
\usepackage{xcolor}
\definecolor{xiaoranpurple}{HTML}{7B1FA2}

\usepackage{etoolbox}
\makeatletter
\patchcmd{\@maketitle}{\vskip21pt}{\vskip0pt}{}{\PackageError{v21}{Title spacing patch failed}{}}
\patchcmd{\@maketitle}{\vskip20pt}{\vskip10pt}{}{\PackageError{v21}{Title spacing patch failed}{}}
\patchcmd{\@maketitle}{\vskip20pt}{\vskip12pt}{}{\PackageError{v21}{Author spacing patch failed}{}}
\patchcmd{\@maketitle}{\vskip24pt}{\vskip10pt}{}{\PackageError{v21}{Address spacing patch failed}{}}
\patchcmd{\@maketitle}{\vskip24pt}{\vskip10pt}{}{\PackageError{v21}{Abstract spacing patch failed}{}}
\makeatother
\usepackage{textcomp}
\usepackage{manyfoot}
\usepackage{booktabs}                          % 三线表
\usepackage{multirow}
\usepackage[normalem]{ulem}
\begin{document}

%=======================================================%
%  标题: 方括号内为页眉用短标题，大括号内为完整标题
%=======================================================%

\title[]{Experimental demonstration of broadband-laser suppression of cross-beam energy transfer}

%=======================================================%
%  作者与单位
%  * 号标记通讯作者; \fnm{名} \sur{姓};
%  作者后的 [1,2] 为单位编号，与下方 \affil 对应;
%  并列第一作者用 \equalcont{...} 声明
%=======================================================%
% ==================== 作者（* = 通讯作者） ====================

\author[1]{\fnm{Guoxiao} \sur{Xu}}
\equalcont{These authors contributed equally to this work.}

\author[2]{\fnm{Xiaoran} \sur{Li}}
\equalcont{These authors contributed equally to this work.}

\author*[3]{\fnm{Ning} \sur{Kang}}\email{kangning@siom.ac.cn}

\author[3]{\fnm{Huiya} \sur{Liu}}

\author*[2]{\fnm{Liang} \sur{Hao}}\email{hao\_liang@iapcm.ac.cn}

\author[2]{\fnm{Zhiyuan} \sur{Li}}

\author[2]{\fnm{Guowei} \sur{Yang}}

\author[1]{\fnm{Honghai} \sur{An}}

\author[1]{\fnm{Xichen} \sur{Zhou}}

\author[1]{\fnm{Jian} \sur{Wang}}

\author[1]{\fnm{Lin} \sur{Yi}}

\author[1]{\fnm{Jun} \sur{Xiong}}

\author[1]{\fnm{Zhiyong} \sur{Xie}}

\author[1]{\fnm{Junjian} \sur{Ye}}

\author[1]{\fnm{Yuchun} \sur{Tu}}

\author[1]{\fnm{Zhiheng} \sur{Fang}}

\author[1]{\fnm{Guo} \sur{Jia}}

\author[1]{\fnm{Wei} \sur{Wang}}

\author[1]{\fnm{Lan} \sur{Xia}}

\author[1]{\fnm{Wei} \sur{Feng}}

\author[1]{\fnm{Lailin} \sur{Ji}}

\author[1]{\fnm{Xiaohui} \sur{Zhao}}

\author*[1]{\fnm{Anle} \sur{Lei}}\email{lal\_leianle@163.com}

\author*[2]{\fnm{Lifeng} \sur{Wang}}\email{wang\_lifeng@iapcm.ac.cn}

\author[1]{\fnm{Yanqi} \sur{Gao}}

% ==================== 单位 ====================
\affil*[1]{\orgname{Shanghai Institute of Laser Plasma, China Academy of Engineering Physics}, \orgaddress{\city{Shanghai}, \postcode{201800}, \country{China}}}

\affil*[2]{\orgname{Institute of Applied Physics and Computational Mathematics}, \orgaddress{\city{Beijing}, \postcode{100094}, \country{China}}}

\affil*[3]{\orgname{Joint Laboratory of High Power Laser and Physics, Shanghai Institute of Optics and Fine Mechanics, Chinese Academy of Sciences}, \orgaddress{\city{Shanghai}, \postcode{201800}, \country{China}}}

%=======================================================%
%  摘要 (Abstract)
%  Nature 期刊要求: 无结构摘要，通常不超过约 150 词，
%  不含子标题、公式和文献引用; 缩写首次出现时展开。
%  下面为示例草稿（约 230 词），投稿前请按期刊要求压缩，
%  并确认 "CBET" 等缩写是否需要在摘要中展开。
%=======================================================%

\abstract{
	In direct-drive inertial confinement fusion (ICF), cross-beam energy transfer (CBET) redirects a significant fraction of the incident laser energy out of the plasma.
	We report the experimental demonstration that broadband lasers reduce CBET-enhanced reflected-light return, performed at the low-coherence Kunwu laser facility with two crossed beams of 0.6\% bandwidth and up to 550\,J at $\sim$2.6$\times$10$^{14}$\,W/cm$^{2}$.
	A coupled ray-tracing model shows that the strong CBET amplification of narrowband reflected light is much weaker under broadband illumination.
	In symmetric incidence condition of two orthogonal beams, the total fractional scattered energy decreases from 7.57\% to 4.64\%; in asymmetric incidence condition, which separates stimulated Brillouin scattering (SBS) from specular reflection, the SBS fraction decreases from 3.75\% to 0.46\% and the specular-reflection fraction decreases from approximately 2.2\% to 1.3\%.
	These results establish broadband lasers as an effective, experimentally validated approach to reducing energy escape through CBET-enhanced reflection in direct-drive ICF.}

% 关键词（Nature 期刊一般不需要，如编辑要求再启用）
% \keywords{keyword1, keyword2, keyword3}

\maketitle

%=======================================================%
%  正文
%  Nature 类论文典型结构: Introduction → Results → Discussion
%  Results 内允许使用子标题; Introduction 不要使用子标题
%=======================================================%

%\section{Introduction}\label{sec:intro}

Laser direct-drive ICF provides a pathway to high-gain thermonuclear burn for inertial fusion energy \cite{betti2016,williams2024}.
Implosion performance is influenced by ablation pressure, which depends on the effective deposition of laser energy into the target.
Ablation produces a low-density plasma corona where a variety of laser plasma instabilities (LPI) occur \cite{randall1981,seka2002}.
These instabilities reduce laser absorption and cause preheat limiting energy gain.
Among them, cross-beam energy transfer (CBET) is a dominant LPI process leading to laser energy loss \cite{edgell2017,froula2025}.
CBET occurs when a pump beam and a seed beam interfere, creating a beat frequency that generates resonant ion acoustic waves and modulates plasma density.
This modulation changes the direction of laser transmission through Bragg diffraction \cite{michel2009,hao2025,debayle2025,oudin2025,longman2025}, resulting in reduced beam--target coupling efficiency and diminished hydrodynamic efficiency during target compression \cite{igumenshchev2012,marozas2018}.
Early studies estimate that CBET causes a laser energy loss of 20\%--40\% relative to the no-CBET limit \cite{froula2013}, making CBET mitigation essential for achieving the high ablation pressures required in high-gain direct-drive ICF \cite{betti2016,trickey2024,goncharov2024}.

CBET mitigation has relied on two approaches, reducing the beam-to-target radius and implementing beam-to-beam frequency detuning \cite{igumenshchev2012,marozas2018,froula2013,anderson2024,oudin2021}.
By reducing beam size, one can eliminate edge-beam rays that seed CBET \cite{igumenshchev2012}.
The three-leg geometry of OMEGA theoretically demonstrates that a wavelength detuning of 2 nm for 351-nm lasers (0.57\%) can effectively reduce CBET while concurrently improving irradiation uniformity \cite{edgell2017}.
Polar-direct-drive experiments with 0.13\% wavelength detuning at the National Ignition Facility show a 16\% increase in equatorial implosion velocity, consistent with radiation-hydrodynamic predictions of a 10\% increase in the average ablation pressure \cite{marozas2018}, corroborating the effect of wavelength detuning on CBET.
Nevertheless, the gains achieved by these approaches remain below those required for ignition-scale direct-drive designs, underscoring the need for additional mitigation techniques.

Low-coherence lasers, which possess a continuous broadband spectrum, have been proved effective in mitigating stimulated Brillouin scattering (SBS) by single-beam experiments with 0.5\%--0.6\% bandwidth \cite{lei2024,kang2025,kanstein2025}.
A substantial number of simulations have been carried out to investigate the mitigation of CBET through continuous bandwidth, indicating a bandwidth of $\sim$1\% of 351 nm wavelength may suppress CBET to a negligible level \cite{bates2018,seaton2022,zhao2023,bates2023}.
Currently, the efficacy of broadband-laser suppression of CBET has not been experimentally demonstrated, mainly owing to the energy deficiency of broadband drivers \cite{major2024,dorrer2026}, which restricts practical experiments to single‑beam setups.

In this work, we report dual-beam evidence for reduced CBET-enhanced reflected-light return under broadband illumination at the Kunwu low-coherence laser facility.
Both simulations and measurements demonstrate a clear reduction in energy loss due to CBET-enhanced reflection and SBS backscatter.
Our results establish broadband laser drive as a practical means of reducing energy escape associated with CBET-enhanced reflection, providing a path toward the high ablation pressures required for ignition-scale direct-drive targets.

%\section{Results}\label{sec:results}

\begin{figure}[!b]
	\centering
	\includegraphics[width=0.85\textwidth]{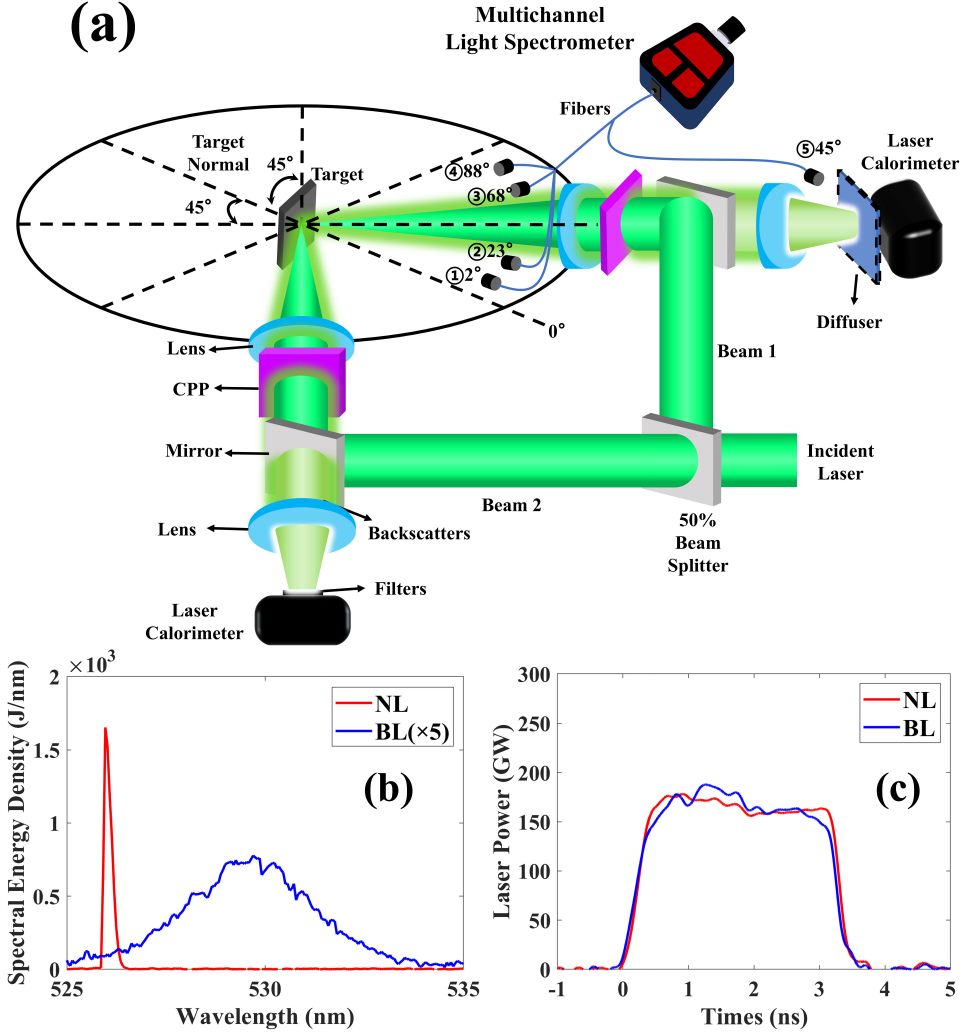}
	\caption{(a) Configuration of the experiment platform for the symmetric case. The target is rotated for the asymmetric case while the beam and diagnostic directions remain fixed. (b) Narrowband and broadband laser spectra at $\sim$550\,J energy. The central wavelengths are 526.1 and 529.5 nm, respectively; the fractional bandwidths are below 0.01\% and approximately 0.6\%. The broadband spectrum is multiplied by five for display. (c) Waveforms of narrowband and broadband laser pulses.}
	\label{setup}
\end{figure}

Figure~\ref{setup} summarizes the experimental platform and laser conditions.
The Kunwu laser is split equally into two orthogonal beams that irradiate an 80\,$\upmu$m-thick planar polystyrene (CH) target with up to $\sim$550\,J total energy.
Both beams are polarized perpendicular to the plane containing their axes and focused at $f$/5.3.
We compare narrowband illumination with a fractional bandwidth below 0.01\% and broadband illumination with a 0.6\% bandwidth (full width at half maximum).
The two modes are produced by frequency doubling a conventional neodymium-glass laser pulse and an amplified superluminescent pulse, respectively.
Identical continuous phase plates (CPPs) produce 210\,$\upmu$m-diameter spots in the plane perpendicular to each beam axis.
Both modes have 3.2-ns square pulse waveforms, giving an average intensity of $\sim$2.6$\times$10$^{14}$\,W/cm$^{2}$ per beam.

In order to comprehensively verify the suppression effect of broadband laser, we designed two cases of laser incidence angles.
In the symmetric case, both beams are incident at 45$^\circ$ to the target normal, and each entrance-port diagnostic receives SBS from its incident beam together with reflected light from the opposite beam.
Calibrated calorimeters measure the return at both entrance ports, while spectra near the laser wavelength are recorded at the Beam 1 entrance port and at 2$^\circ$, 23$^\circ$, 68$^\circ$, and 88$^\circ$ around it, as shown in Fig.~\ref{setup}.
These diagnostic angles are referenced to the bisector of the incident beams.
In the asymmetric case, the target is rotated counterclockwise by approximately 12$^\circ$ giving nominal incidence angles of 33$^\circ$ and 57$^\circ$, while the laser entrance-port and diagnostic ports are kept at the same position.
Thus the principal specular-reflection directions are separated from the entrance ports, allowing SBS and reflected light to be diagnosed separately.
The 68$^\circ$ diagnostic samples mirror reflection of the nominally 57$^\circ$ beam. 

We perform two-dimensional ray-tracing calculations using plasma density, temperature, and flow profiles from radiation-hydrodynamic simulations.
Laser propagation and refraction are solved together with absorption, CBET, and SBS.
The same calibrated model parameters are used for the symmetric and asymmetric cases under narrowband and broadband illumination, without a separate broadband fit.
Model implementation, calibration, and diagnostic definitions are given in Methods.

\begin{figure}[!t]
	\centering
	\includegraphics[width=0.9\textwidth]{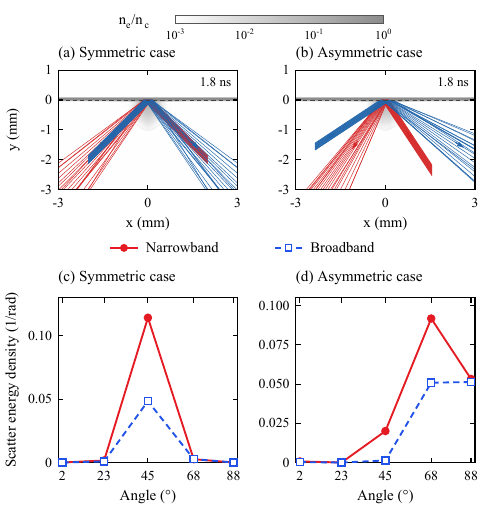}
	\caption{Simulated propagation and angular return for the symmetric (left) and asymmetric (right) cases. (a),(b) Narrowband rays at 1.8 ns; red and blue identify the beams, arrows indicate propagation, gray shading gives $n_{\rm e}/n_{\rm c}$, and the dashed line marks the initial target surface. (c),(d) Narrowband and broadband angular energy densities, integrated over 0.4--3.2 ns within a 10$^\circ$ half-angle and normalized to the total incident energy and aperture width.}
	\label{design}
\end{figure}

Figures~\ref{design}(a) and \ref{design}(b) show ray trajectories at 1.8\,ns for the symmetric and asymmetric cases, respectively.
In the symmetric case, each entrance port collects SBS backscatter from its incident beam together with reflected light from the opposite beam.
In the asymmetric case, reflection of the nominally 57$^\circ$ beam is directed toward the 68$^\circ$ diagnostic window, spatially separated from the entrance ports.
Figures~\ref{design}(c) and \ref{design}(d) give the simulated angular energy densities at the diagnostic angles shown in Fig.~\ref{setup}.
The principal narrowband peaks occur at 45$^\circ$ and 68$^\circ$ in the symmetric and asymmetric cases, respectively, and the signals at both angles decrease markedly under broadband illumination.
For the symmetric case, the energy collected at 45$^\circ$, normalized to the incident energy of one beam, falls from 7.98\% to 3.38\%.
This mixed return can decrease through both weaker SBS backscatter and reduced CBET amplification of reflected light.
Both processes involve resonant coupling through ion-acoustic waves, and spectral broadening lowers their effective gain in the present model.
For the asymmetric case, the 45$^\circ$ entrance-port fraction, normalized to the incident energy of the 33$^\circ$ beam, falls from 1.40\% to 0.09\%, with the decrease dominated by SBS suppression.
Meanwhile, the fraction in the 68$^\circ$ reflection window, normalized to the incident energy of the 57$^\circ$ beam, falls from 6.41\% to 3.55\%, a reduction of about 45\%.
This decrease is primarily associated with weaker CBET amplification of reflected light, as assessed by the paired calculations below.
Bandwidth reduces the spectrum-averaged inter-beam coupling \cite{follett2023}, limiting the energy gained by the reflected branch of the 57$^\circ$ beam from the incoming 33$^\circ$ beam.

\begin{figure}[!t]
	\centering
	\includegraphics[width=0.9\textwidth]{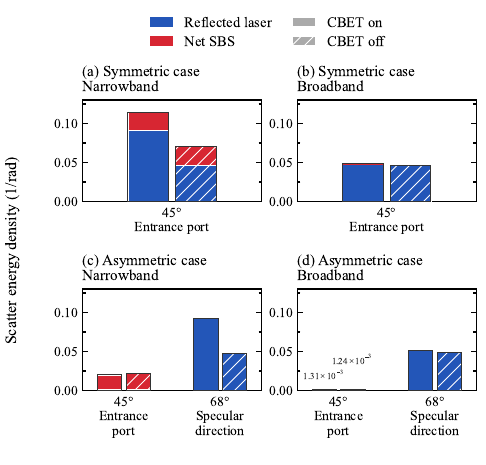}
	\caption{Simulated received energy with CBET enabled (solid) and disabled (hatched), resolved into reflected laser light (blue) and net SBS (red). Each pair uses the same background and other parameters, with the diagnostics of Fig.~\ref{design}. Symmetric incidence superposes the components at 45$^\circ$; asymmetric incidence separates the entrance port at 45$^\circ$ from mirror reflection at 68$^\circ$.}
	\label{simulation}
\end{figure}

To identify the role of inter-beam coupling, we repeat each calculation with CBET disabled, holding the plasma background, input beams, absorption and SBS treatment, and other parameters fixed as shown in Fig.~\ref{simulation}.
The received light is resolved into reflected laser light and net SBS.
The on/off difference includes power redistribution and its feedback on SBS and absorption; it is a response of the coupled system, not an independently additive optical component.
For narrowband illumination, disabling CBET lowers the signal in the principal reflection direction by 38\% in the symmetric case and 48\% in the asymmetric case.
The decrease is dominated by reflected laser light.
In contrast, the SBS-dominated signal at the asymmetric entrance port increases by about 9\%.
CBET therefore enhances escape mainly by amplifying reflected light, with a different response in the SBS channel.
For broadband illumination, disabling CBET changes the signals at these reflection directions by only 3\%--4\%, while the SBS component is strongly reduced relative to narrowband illumination.
The calculations therefore identify two effects to test experimentally: suppression of SBS at the asymmetric entrance ports and weaker CBET enhancement of light escaping in the separated reflection direction.
Both contributions enter the symmetric port return.

The entrance-port measurements show the predicted reduction in both geometries.
In the symmetric case, three shots at each bandwidth give arithmetic mean two-port fractions of 7.57\% for narrowband illumination and 4.64\% for broadband illumination, a reduction of 39\% between the means.
Each shot is normalized to its total incident energy before averaging.
The individual-shot ranges are 7.42\%--7.72\% and 4.50\%--4.89\%, respectively.
This decrease in combined SBS and reflected light is consistent with the calculated change from 7.98\% to 3.38\%, although the model underestimates the broadband return.
The asymmetric entrance ports primarily diagnose SBS because the principal reflected light leaves in other directions.
Two shots at each bandwidth give a mean two-port fraction of 3.75\% for narrowband illumination and 0.46\% for broadband illumination, a reduction of 88\%.
The corresponding individual-shot ranges are 3.67\%--3.84\% and 0.44\%--0.49\%.
These measurements directly demonstrate substantial SBS suppression under dual-beam broadband irradiation.
The separated reflection direction tests whether the bandwidth effect extends beyond this SBS channel.

\begin{figure}[!b]
	\centering
	\includegraphics[width=0.9\textwidth]{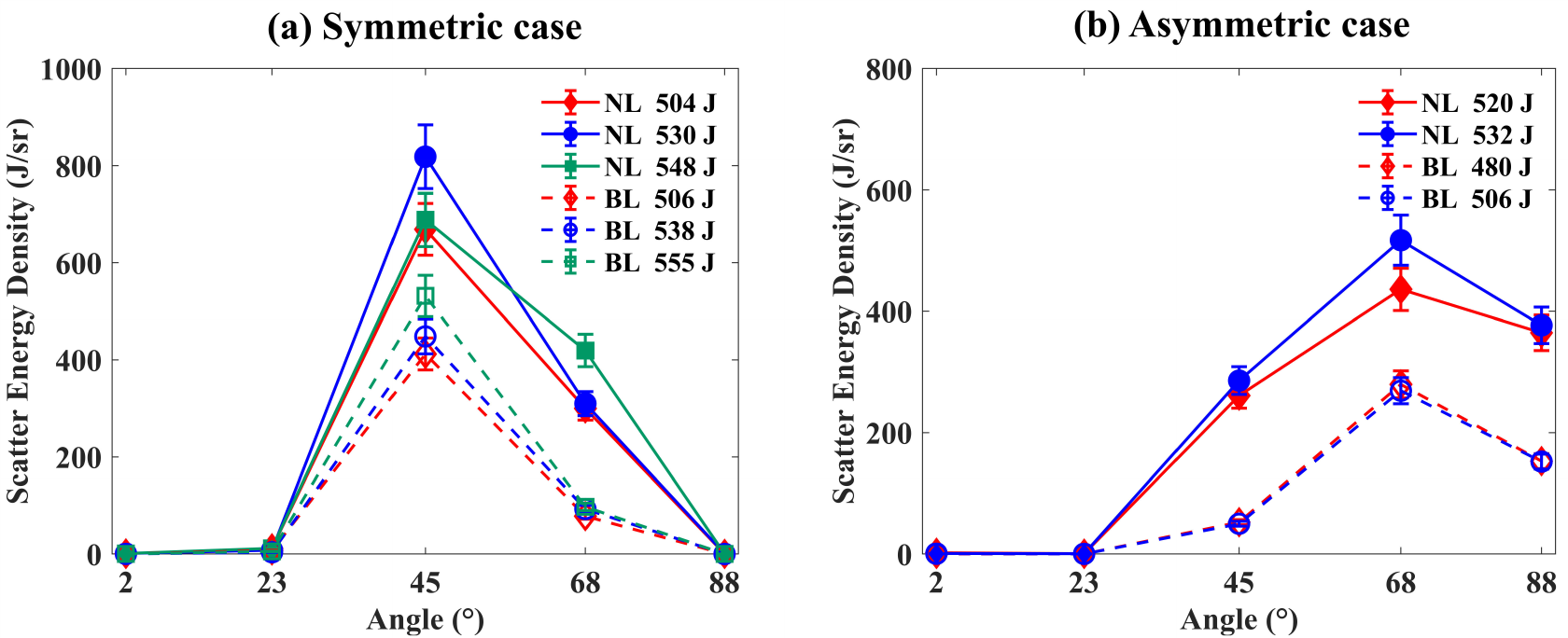}
	\caption{Measured angular energy densities for dual-beam shots in the (a) symmetric and (b) asymmetric cases. Solid and dashed curves denote narrowband (NL) and broadband (BL) illumination. The legends list the incident energies of the individual shots. Angles are referenced to the beam bisector: 45$^\circ$ is an entrance port, and 68$^\circ$ samples the principal mirror reflection in the asymmetric case.}
	\label{measurement}
\end{figure}

Figure~\ref{measurement} compares the measured angular energy densities.
In the symmetric case [Fig.~\ref{measurement}(a)], the signal is strongest at 45$^\circ$, followed by 68$^\circ$, and much weaker at the other angles.
Broadband signals are lower at all measured directions.
The dominant entrance-port signal is consistent with the overlap of SBS and reflected light shown by the simulated trajectories and component decomposition.
In the asymmetric case [Fig.~\ref{measurement}(b)], the dominant signal shifts to 68$^\circ$, the mirror direction of the nominally 57$^\circ$ beam.
Its collected fraction decreases from approximately 2.2\% to 1.3\% of the total incident energy, a reduction of about 41\%.
Thus, bandwidth reduces the spatially separated reflected-light channel as well as the SBS-dominated entrance-port return.

The measured shift of the principal signal follows the reflection geometry predicted by the model.
At the asymmetric mirror direction, the calculated reduction of about 45\% is comparable to the measured 41\%, although the absolute fractions differ.
Reflection alone does not isolate CBET from propagation and absorption.
Combined with the switch calculations, however, its decrease supports suppression of CBET-enhanced escape in addition to SBS backscatter.

%\section{Discussion}\label{sec:discussion}

Although the experimental measurement results are generally in line with theoretical expectations, some quantitative differences remain between the simulations and measurements.
The ray-tracing model especially overestimates broadband SBS suppression.
Temporal intensity fluctuations absent from the fixed-spectrum treatment \cite{follett2023} can produce high-intensity intervals that drive SBS \cite{li2026} and may contribute to this discrepancy.
These limitations preclude a precise partition of the measured return, but do not remove the distinct channel responses identified by the paired calculations.
The comparison concerns finite diagnostic apertures in prescribed plasma backgrounds.
It does not determine total escaping energy or absorption from the measured return alone, and the calculations exclude hydrodynamic feedback.
Within this scope, the combined evidence links lower return to reduced SBS and weaker CBET amplification of reflected light.

In summary, CBET has long posed a fundamental obstacle to high-performance direct-drive ICF, limiting the ablation pressure.
Our dual-beam experiments at the Kunwu facility, interpreted with coupled ray-tracing simulations, indicate that broadband lasers reduce CBET-enhanced reflected-light return as well as SBS backscatter.
The measured reduction, predicted by a coupled ray-tracing model, supports the interpretation that CBET is a significant contributor to energy loss via its redistribution of laser energy into reflected light, and indicates that this pathway is substantially weakened by the introduction of laser bandwidth.
By spatially separating the principal SBS and specular-reflection directions through asymmetric incidence, the experiments show reduced return in both channels under broadband illumination.
These findings establish broadband laser drive as a means of reducing energy escape through CBET-enhanced reflection, advancing the prospects for the high ablation pressures required in ignition-scale direct-drive designs.
The low-coherence approach demonstrated here, based on frequency doubling an amplified superluminescent pulse, is compatible with existing high-energy laser architectures, offering a practical route toward broadband drivers for inertial fusion energy.

%=======================================================%
%  方法 (Methods)
%  Nature 期刊: Methods 放在参考文献之前;
%  需包含他人可复现实验所需的充分信息
%=======================================================%

\section*{Methods}

\subsection*{Experimental platform and diagnostic calibration}
The Kunwu laser is split by a 50\% beam splitter.
The two beams irradiate the target orthogonally at the chamber center, with their polarizations perpendicular to the horizontal plane defined by the beam axes.
The $f$-number of each beam is 5.3.
In the narrowband mode, the laser is the second harmonic of a conventional neodymium glass laser.
In the broadband mode, an amplified superluminescent pulse from a neodymium glass amplifier is frequency-doubled to produce green broadband light.
The central wavelengths are 526.1 and 529.5\,nm, respectively.

The backscatter from each beam is collected by a lens and measured with a calibrated laser calorimeter.
Spectral measurement of Beam 1 backscatter is available when a diffuser is placed in front of the calorimeter, which blocks the energy measurement.
A band-pass filter centered at the laser wavelength is placed in front of the calorimeters to reject Raman scattering.
The spectrometer is centered at the laser wavelength to cover the range of 525--535\,nm.
The spectrometer uses a 2400-grooves/mm grating with a spectral resolution of 0.05\,nm.
For absolute calibration of the energy measurement, a continuous-wave laser with the same wavelength, polarization, path, and aperture as the Kunwu laser is used.
The spectral diagnostic is calibrated with a standard light source of known spectral power distribution.

\subsection*{Ray propagation and coupled energy exchange}
We perform two-dimensional ray-tracing calculations using plasma density, temperature, and flow profiles from radiation-hydrodynamic simulations.
Laser propagation and refraction are solved together with absorption, CBET, and SBS during the pulse plateau.
The incident beams refract and turn in the inhomogeneous plasma while their powers evolve through absorption and wave-mediated energy exchange, determining the reflected and backscattered light received by each diagnostic.
Each beam is represented by 20 spatial rays with a ray-path step of 0.8\,$\upmu$m.
The asymmetric calculation uses incidence angles of 33.5$^\circ$ and 56.5$^\circ$, corresponding to the rounded experimental angles.
Each plasma state is prescribed; the propagation calculation does not update the hydrodynamic background.

\subsection*{Broadband treatment}
Broadband CBET is treated with the fixed-spectrum approximation \cite{follett2023}, averaging the coupling coefficient over frequency pairs weighted by the prescribed top-hat spectra. This average uses ten equally weighted spectral points per beam spanning a full bandwidth of 0.6\% about 529.5\,nm.
The resulting coefficient governs energy exchange using local beam powers, while the relative spectral weights remain fixed.
The SBS calculation retains the frequency-dependent gain for backward Stokes light along reversed pump-ray paths.
This treatment captures spectral changes in resonant coupling but does not resolve phase interference or rapid temporal intensity fluctuations.

\subsection*{Phenomenological parameters and single-beam calibration}
An upper limit on the acoustic density perturbation represents ion-acoustic saturation.
Because CPP speckles are not explicitly resolved, an effective SBS gain multiplier accounts phenomenologically for their intensity modulation.
These effective parameters describe unresolved physics within the ray model; they are not independent measurements of the microscopic saturation mechanism.

In addition to the four dual-beam configurations discussed in the main text, Beam-1-only and Beam-2-only shots are obtained by blocking one path after the beam splitter, while both return diagnostics remain available.
In single-beam shots at 45$^\circ$ incidence, SBS backscatter and specular reflection are measured in different directions, providing separate constraints on the model.
The single-beam 45$^\circ$ measurements give narrowband SBS backscatter of 0.92\%--1.62\% and specular reflection of 1.83\%--2.03\%, each relative to the single-beam incident energy.
The phenomenological parameters used in our model were constrained by the single-beam shots, and the chosen parameters were held fixed across the four dual-beam configurations.

The single-beam broadband measurements retain 0.14\%--0.20\% SBS backscatter, relative to the single-beam incident energy.
This residual return provides an additional indication that a spectrum-averaged treatment need not reproduce the full broadband SBS response.
Rapid temporal intensity fluctuations, absent from the fixed-spectrum calculation, may contribute to the residual SBS \cite{li2026}.

\subsection*{Synthetic diagnostics and CBET control calculations}
For the dual-beam comparisons, simulated light is integrated over 0.4--3.2\,ns within a 10$^\circ$ half-angle about each experimental diagnostic direction.
Fifteen plasma states, separated by 0.2\,ns, sample this interval.
Fractions are normalized to the total incident energy over that interval; division by the aperture width gives the angular density in 1/rad.
Individual-window energy fractions in the discussion of Fig.~\ref{design} are normalized to the corresponding beam energy; the plotted angular densities and two-port fractions use the total incident energy of both beams.
The plateau calculation approximates the full-pulse experimental response.
Its two-dimensional density and the measured density in J/sr are compared through angular features and integrated-fraction trends, rather than as identical absolute quantities.

The same diagnostic apertures and integration interval are used in every paired CBET-on/off calculation.
Each pair holds the plasma background, input beams, absorption and SBS treatment, and other parameters fixed.
Received light is resolved into reflected laser light and net SBS; net SBS is obtained by subtracting the propagated seed-only baseline.
The resulting difference between calculations includes the feedback of inter-beam coupling on SBS and absorption.

\begingroup
\let\vxxioriginalbibitem\bibitem
\renewcommand{\bibitem}[1]{%
%  \ifstrequal{#1}{trickey2024}{\color{red}}{%
%    \ifstrequal{#1}{goncharov2024}{\color{red}}{\color{black}}}%
  \vxxioriginalbibitem{#1}}
\bibliography{sn-bibliography}
\endgroup

% eJP 投稿前: 将编译生成的 template.bbl 内容粘贴到此处，
% 并删除上面 \bibliography{sn-bibliography} 一行。

%=======================================================%
%  文末声明 (backmatter)
%  Nature 期刊顺序: Data availability → Code availability
%  → Acknowledgements → Author contributions → Competing interests
%=======================================================%
\backmatter

\bmhead{Data availability}
The data that support the findings of this study are available from the corresponding authors upon reasonable request.

\bmhead{Code availability}
% 如有可公开代码请注明仓库与版本; 不适用则写 Not applicable
The simulation codes used in this manuscript are not available to the 
general public.

\bmhead{Acknowledgements}
% 致谢与资助信息，注明资助号（如有）
We gratefully acknowledge the beneficial assistance and help of all the technical staff during the experiments.
This work was supported by the National Natural Science Foundation of China (Grant Nos. 12588301, 11905280, 12635016, 12275032, 12205021, 12505268) and the National Key R\&D Program of China (Grant No. 2023YFA1608400).

\section*{Author contributions}
% 逐条列出各作者贡献，例如:
% F.A. conceived the experiments. S.A. performed the experiments and analysed the data. All authors discussed the results and contributed to the manuscript.
These authors contributed equally: G.X. and X.L.
N.K., L.H., A.L., and L.W. conceived the idea.
G.X. and X.L. wrote the paper.
N.K. and A.L. led the experimental arrangement, diagnostic plan and measurement synthesis, with the help from G.X. and H.L.
L.H. and L.W. coordinated the theoretical design and analysis of the experiment, with the help from X.L.
Z.L. and G.Y. performed the hydrodynamic simulations.
H.A., X.Z., J.W., L.Y., and J.X. operated diagnostics.
Z.X., J.Y., and Y.T. fabricated targets.
L.X., W.F., L.J., X.Z., and Y.G. established the dual-beam platform and operated the laser.
Z.F., G.J., and W.W. participated in some discussions.

\section*{Competing interests}
The authors declare no competing interests.

%=======================================================%
%  Extended Data
%  Nature 期刊使用 "Extended Data" 而非附录; 投稿时
%  Extended Data 图表作为单独文件上传，正文中以
%  "Extended Data Fig. 1" 等引用。如需在 tex 中起草:
%=======================================================%
% \begin{appendices}
% \section{Extended Data Fig. 1 ...}\label{secA1}
% \end{appendices}

\end{document}